# Fluctuations in the Gravitational, Strong and Weak Nuclear Fields through an Effective Harmonic Oscillator Model.

P. R. SILVA

DEPARTAMENTO DE FÍSICA - INSTITUTO DE CIÊNCIAS EXATAS

UNIVERSIDADE FEDERAL DE MINAS GERAIS

CAIXA POSTAL 702

30.123-970 Belo Horizonte - MG - Brazil

## Abstract

We propose an effective harmonic oscillator model in order to treat the fluctuations of the gravitational, strong and weak nuclear fields. With respect to the gravitational field, first we use the model to estimate its fluctuating strength, necessary to decohere the wavefunction of a cubic centimeter of air at the standard temperature and pressure conditions. Second, the fluctuation of a point mass through a distance equal to the Planck length leads to the self-gravitational interaction of a particle, which can be related to its de Broglie frequency. Third, by making the equality of the fluctuating field strength with the gravitational field of a mass M at half of its Schwarzschild radius, we obtain an estimate of the mass of the Universe. We also consider the fluctuations of the strong nuclear field, as a means to estimate the separation in energy between the ground state and the centroid of the excited states of the nucleon. Finally, taking into account the neutron-proton mass difference, we use the fluctuations of the weak nuclear field in order to evaluate the weak coupling constant.

**1- Introduction**

In his paper: "Why do we observe a classical spacetime?", Joos [1] concludes that the smallness of the Planck's length is not sufficient to explain the absence of quantum effects of gravitation. As was pointed out by Joos [1]: "Instead, in order to become classical, spacetime has to be continuously measured by matter." In order to demonstrate this last statement Joos has considered essentially the fluctuations of the gravitational field, when measured by the content of a cubic centimeter of air at the standard temperature and pressure conditions (STP), during a period of one second.

In this letter, we intend to show that, working in a similar way as Joos did, we can get information about the structure of matter, perhaps of far reaching consequences.

In section 2, through an oversimplified calculation, we evaluate a relative fluctuation of the gravitational field which is two orders of magnitude smaller than that obtained through Joos more refined calculations [1]. In section 3 we use the same method as a means of speculating about the origin of the driving force responsible for the establishment of the de Broglie frequency. In the section 4, the present "formalism" is used to estimate the mass of the Universe. Section 5 applies this "formalism" to the fluctuations of the color field. The estimation of the weak force strength is worked out in section 6. We finish this letter in section 7 with some concluding remarks.

**2- An alternative way to Joos derivation**

Let us consider a typical particle of rest mass m coupled to a fluctuating local gravitational field $\Delta g_i$. We suppose that, under the action of this field, this particle suffers a displacement $x_i$ from its mean position. If we assume, as it seems to be natural, that the fluctuating field has equal probability to take negative and positive values, the first non-zero contribution for the interaction between the particle and the field comes from the second moment of this interaction. So, we can write for N particles

$$\Delta U = \frac{N^2 m^2 (\Delta g)^2 x^2}{E_0}. \qquad (1)$$

In (1), $\Delta g$ and x are the moduli of the fluctuating field and of the displacement respectively and $E_0$ is a normalizing factor, so that $\Delta U$ will have the dimension of energy. Choosing the sample of 1 cm$^3$ of air as the tool used to measure the fluctuating field, we suppose $E_0$ to be equal to the mean kinetic energy of the air particles, summed over the N particles present in the cited volume of air at the STP. Then, we can write

$$\frac{N^2 m^2 (\Delta g)^2 x^2}{\frac{1}{2} N m v_s^2} = \frac{1}{2} k x^2 . \qquad (2)$$

In (2), we identify the effective potential $\Delta U$ with a harmonic oscillator of spring constant k and $v_s$ stands for the velocity of the sound in air at the STP.

Then, the angular frequency $\omega$ is given by

$$\omega = \sqrt{\frac{k}{m}} = \frac{2\sqrt{N}\Delta g}{v_s} . \qquad (3)$$

The maximum variation of the phase of the wave function of this effective quantum oscillator will occur when

$$\omega t = 2\pi .$$

Using this result in (3) for t=1s, $\sqrt{N} \sim 5 \times 10^9$ and $v_s \approx 3.5 \times 10^2$ m/s, we obtain $\Delta g \sim 10^{-7}$ m/s$^2$ and

$$\Delta g/g \sim 10^{-8}.$$

The relative fluctuation of the gravitational field evaluated here is two orders of magnitude smaller than the result obtained by Joos [1]. However, we can extract some other interesting results if we consider other kinds of fields and even the gravitational field acting in other circunstances, working in an analogous way as that presented in this section. This is the objective of the following sections.

### 3- Origin of the de Broglie frequency

Let us consider a single particle (N=1). Here, in order to normalize the effective interaction of the fluctuating gravitational field, we will take the "normalizing" energy $E_0 = 2mc^2$, when the field strength is sufficiently high to create a pair of particles of rest mass m. Then, we can write:

$$\frac{m^2 (\Delta g)^2 x^2}{2mc^2} = \frac{1}{2} k_I x^2. \tag{4}$$

The angular frequency of the harmonic oscillator, which spring constant is given by (4), is

$$\omega_I = \sqrt{\frac{k_I}{m}} = \frac{\Delta g}{c}. \tag{5}$$

Penrose [2] attributes the existence of accurate clocks to the possibility of associating a de Broglie frequency [3] to each particle of mass m. Also, as was pointed out by Anandan [4], from an operational point of view, gravity appears to be deeply rooted in the wave particle duality of matter. Then it seems that to identify the frequency due to a fluctuating local gravitational field, given by (5), with the de Broglie frequency, could be perhaps a reasonable hyphothesis. Taking into account these considerations, we can write

$$\hbar \omega_I = \hbar \frac{\Delta g}{c} = mc^2, \tag{6}$$

which implies,

$$\Delta g = \frac{Gm}{\lambda_P^2}, \tag{7}$$

where $\lambda_P$ is the Planck's length given by $\lambda_P^2 = \left(\frac{G\hbar}{c^3}\right)$, being G the gravitational constant, $\hbar$ the reduced Planck's constant and c the speed of light. Then, we can imagine that the mass of a point particle could fluctuate in position through a distance equal to the Planck length. This generates a fluctuating local gravitational field and the coupling of the particle to this field (self-interaction) could be interpreted as the driving force responsible for the particle "internal motion" which sets up the Penrose [2,4] clock. A result similar to (7) has been obtained by the present author [5], through somewhat different reasonings.

### 4- The Mass of the Universe

We can also compare the $\Delta g$ given by (5) with the gravitational field of the Universe of mass M, at half of its Schwarzschild radius $r_s$. We have

$$c\omega = \frac{GM}{\left(\frac{1}{2}r_s\right)^2} = \frac{GM}{\left(\frac{GM}{c^2}\right)^2} = \frac{c^4}{GM}. \tag{8}$$

If we indentify $\omega$ with the Hubble constant $H_0$, we obtain

$$M = \frac{c^3}{GH_0}. \tag{9}$$

The above result is the same as that found in a paper by Sharma and Sharma [6] as an upper bound on the particle mass, which they proposed to be equal to the mass of the Universe.

### 5- Fluctuations of the Color Field and the Nucleon energy levels

We can treat the field associated to the strong nuclear force in an analogous way whith that we have done before, in the gravitational interaction case. So, we can write

$$\frac{e_s^2(\Delta\Psi)^2 x^2}{2m_\pi c^2} = \frac{1}{2}k_s x^2. \tag{10}$$

In (10), $\Delta\Psi$ is the fluctuation amplitude of the strong color field and $e_s$ is the color charge. We will take the normalization energy as $E_0=2m_\pi c^2$, the energy sufficient for the creation of a pion pair, where $m_\pi$ is the mass of the pion ($\pm$). The effective harmonic oscillator of spring constant $k_s$, leads to the frequency

$$\omega_s = \frac{e_s \Delta\Psi}{m_\pi c}. \tag{11}$$

If we consider, as Yukawa did [7], that the strong nuclear force is intermediated by the pion, we have

$$\hbar\omega_s = m_\pi c^2. \tag{12}$$

It is worth to notice that, on writing (12), we have identified $\omega_s$ with the de Broglie frequency of the pion. From (11) and (12) we obtain, for the strong field strength,

$$\Delta\Psi = \frac{m_\pi^2 c^3}{e_s \hbar}. \tag{13}$$

It is instructive to write the result (13) in the form

$$e_s \Delta\Psi = \frac{Gm_\pi^2}{\lambda_P^2}. \tag{14}$$

In (14) the left side gives the amplitude of the strong force and the right side is just the amplitude of the driving force associated to the interpretation of the de Broglie frequency of the pion through a harmonic oscillator model [5] (see also (7) and the subsequent discussion).

It would be interesting to make an evaluation of $\Delta\Psi$ in terms of an equivalent electric field. If we consider $e_s \sim \sqrt{137}e$, where $e$ is the proton electric charge, we obtain

$$\Delta\Psi_{\text{equivalent}} \sim 10^{22} \text{ V/m}.$$

The above value can be compared with the electric field strength at the surface of the atom of Uranium [8] which is $\sim 10^{21}$ V/m.

In order to pursue further with our reasoning, and taking the harmonic oscillator model as a paradigm, let us consider the absolute value of the averaged potential energy of the strong force field as half of the "rest" energy of the pion. We then have,

$$\frac{1}{2} m_\pi c^2 = e_s \Delta\Psi\, r_s. \tag{15}$$

Then we obtain for $r_s$, the mean radius of the nucleon, the relation

$$r_s = \frac{m_\pi c^2}{2\, e_s \Delta\Psi}. \tag{16}$$

From the equivalence principle, we can suppose that a fluctuating aceleration $\Delta a$ must satisfy an equation analogous to (5) for $\Delta g$, namely

$$\Delta a = c\omega. \tag{17}$$

By using Newton's law of motion, we can write

$$c\omega_N = \Delta a = \frac{\alpha_s \hbar c}{m_\pi r_s^2}. \tag{18}$$

The right side of (18) is the strong force divided by the pion mass. Using (13), (16) and (18) and solving for $\omega_N$, we finally obtain

$$\hbar\omega_N = 4\, \alpha_s m_\pi c^2. \tag{19}$$

If we take the strong coupling constant $\alpha_s = 1$, we get $\hbar\omega_N \sim 560$ MeV. This value must be compared with 600 MeV, the separation in energy between the ground state and the centroid of the excited states of the nucleon as quoted by Brown and Rho [9].

### 6- Estimation of the Weak Force Strength

In a similar way as we have treated the strong color field, we consider now the field associated to the weak coupling. To to this, we must take into account that the particles which intermediate the weak coupling are the W$\pm$ bosons [10] of mass $m_w$. Then, working in an analogous way as that of the strong field case, we can write

$$\omega_w = \frac{e_w \Delta\Phi}{m_w c}, \qquad (20)$$

where (20) is the equivalent to (11) of the strong field case, $\Delta\Phi$ is the strength of the weak field and $e_w$ is the "weak charge". Ommiting steps analogous to those which go from (12) to (18), with the proper substitution of $m_\pi, e_s, r_s, \Delta\Psi$ and $\alpha_s$ for $m_w, e_w, r_w, \Delta\Phi$ and $\alpha_w$, respectively, we obtain

$$\hbar\omega_{n-p} = 4\alpha_w m_w c^2. \qquad (21)$$

Equation (21) for the weak field is the equivalent to (19) of the strong field. The energy level spacing in (21) can be interpreted as the mass difference between the neutron and the proton. As it is well known, the weak force is responsible for the neutron decay [11]. Then we can write

$$\alpha_w = \frac{\hbar\omega_{n-p}}{4m_w c^2} = \frac{(m_n - m_p)c^2}{4m_w c^2}. \qquad (22)$$

By considering $m_w = 80.5\,\text{GeV}, m_n = 939.6\,\text{MeV}, m_p = 938.3\,\text{MeV}$ [12,13], we obtain

$$\alpha_w = 4.04 \times 10^{-6}.$$

This value for the weak coupling constant is within the values estimated for it as quoted in some textbooks. (See for example, Halzen and Martin [14] ).

### 7 - Concluding Remarks

In this letter we have treated the fields responsible for various kinds of coupling, namely: gravitational, strong and weak nuclear couplings, through the introduction of an effective harmonic oscillator model. The discrepancy we have met between our treatment and the more accurate one of Joos [1], could be due to the simplified way we have treated the many-body effects of the $10^{19}$ particles present in a cubic centimeter of air at STP. So, the results obtained in section 2 of this letter must be considered with caution. However, by using this effective harmonic oscillator "formalism" , we also have been able to obtain interesting information when we applied it to the study of the driving force behind the de Broglie frequency, the energy level separation in the nucleon, the evaluation of the weak coupling constant through the neutron-proton mass difference, and an estimate of the mass of the Universe.

Finally, although the calculations we have perfomed in this letter are somewhat qualitative, perhaps they can be justified if we take into account the generallity of the applications and the intuitive feeling we have achieved about some fundamental physical phenomena.

**Acknowledgment**

We are grateful to Dr. A. V. de Carvalho for reading the manuscript.

REFERENCES

[1] E. Joos, Phys. Lett. A <u>116</u>, 6 (1986); see also references cited therein.

[2] R. Penrose, In Battelle Recontres, Eds. C.M. De Witt and J.A. Wheeler,
    Benjamin, N. York (1968).

[3] L. de Broglie, Ann. Phys. (Paris) <u>2</u>, 10 (1925); see also: J.W. Haslett, Am. J.  Phys. <u>40</u>, 1315 (1972).

[4] J.S. Anandan, in Quantum Theory of Gravitation, Ed. A.R. Marlow, Proc. of
    Simposium held at Loyola University, New Orleans, May 23-26/1979,
    Academic Press (1980).

[5] P.R. Silva, Phys. Essays <u>10</u>, 628(1997); see also: H.C.G. Caldas and P.R.Silva, Apeiron <u>8</u>(1) (2001), gr-qc/9809007.

[6] M.L. Sharma and N.K. Sharma, Am. J. Phys. 52, 264 (1984).

[7] H. Yukawa, Proc. Phys. Math. Soc. Jpn. 17, 48 (1935).

[8] D. Halliday and R. Resnick, Fundamentals of Physics, John Wiley and Sons (1988). Ch. 24

[9] G.E. Brown and M. Rho, Phys. Today 36, 36 (1983), and references cited in this paper.

[10] W.S.C. Williams, Nuclear and Particle Physics, Clarendon Press, Oxford (1991) Ch. 9 and references cited therein.

[11] P. Langacker and A.K. Mann, Phys. Today 42, 22 (1989); see also references cited in this paper.

[12] Particle Data Group, Phys. Lett. B 204, 1 (1988).

[13] R.J. Blin - Stoyle, Nuclear and Particle Physics, Chapman and Hall, London (1991) Ch. 7

[14] F. Halzen and A.D. Martin, Quark and Leptons: An Introductory Course in Modern Particle Physics, John Wiley and Sons (1984).